\documentclass[aps,prl,twocolumn,superscriptaddress]{revtex4}
\usepackage{graphicx}

\begin{document}

\title{Evolution of spin-wave excitations in ferromagnetic metallic
manganites }

\author{F. Ye}
\affiliation{Center for Neutron Scattering,
Oak Ridge National Laboratory, Oak Ridge, Tennessee 37831-6393,
USA}

\author{Pengcheng Dai}
\email{daip@ornl.gov}
\affiliation{Department of Physics and Astronomy,
The University of Tennessee, Knoxville, Tennessee 37996-1200, USA}
\affiliation{Center for Neutron Scattering,
Oak Ridge National Laboratory, Oak Ridge, Tennessee 37831-6393,
USA}

\author{J. A. Fernandez-Baca}
\affiliation{Center for Neutron Scattering,
Oak Ridge National Laboratory, Oak Ridge, Tennessee 37831-6393,
USA}
\affiliation{Department of Physics and Astronomy,
The University of Tennessee, Knoxville, Tennessee 37996-1200, USA}

\author{Hao Sha}
\affiliation{Department of Physics,
Florida International University, 
Miami, Florida 33199,
USA}

\author{J. W.~Lynn}
\affiliation{NIST Center for Neutron Research,
Gaithersburg, Maryland, 20899, USA}

\author{H.~Kawano-Furukawa}
\affiliation{Department of Physics,
Ochanomizu University, Bunkyo-ku, Tokyo 112-8610, Japan
}

\author{Y.~Tomioka}
\affiliation{Correlated Electron Research Center (CERC), 
Tsukuba 305-0046, Japan}

\author{Y. Tokura}
\affiliation{Correlated Electron Research Center (CERC),
Tsukuba 305-0046, Japan}
\affiliation{Department of Applied Physics, University of Tokyo,
Tokyo 113-8656, Japan}

\author{Jiandi Zhang}
\affiliation{Department of Physics,
Florida International University, 
Miami, Florida 33199,
USA}

\date{\today}

\begin{abstract}
Neutron scattering results are presented for spin-wave excitations of
three ferromagnetic metallic $A_{1-x}A^{\prime}_{x}$MnO$_3$
manganites (where $A$ and $A^\prime$ are
rare- and alkaline-earth ions), which when combined with previous work elucidate the
systematics of the interactions as a function of carrier concentration
$x$, on-site disorder, and strength of the lattice distortion.  The
long wavelength spin dynamics show only a very weak dependence across
the series.  The
ratio of fourth to first neighbor exchange ($J_4/J_1$) that controls
the zone boundary magnon softening changes systematically with $x$,
but does not depend on the other parameters.  None of the prevailing
models can account for these behaviors.
\end{abstract}




\maketitle
 
Determining the evolution of the elementary magnetic excitations in
$A_{1-x}A^{\prime}_{x}$MnO$_3$ (where $A$ and $A^\prime$ are rare- and
alkaline-earth ions respectively) is the first step in understanding
the magnetic interactions in these doped perovskite manganites.
According to the conventional double-exchange (DE) mechanism
\cite{zener51}, the motion of charge carriers in the metallic
state of $A_{1-x}A^{\prime}_{x}$MnO$_3$ establishes a ferromagnetic
(FM) interaction between spins on adjacent $\rm Mn^{3+}$ and $\rm
Mn^{4+}$ sites. In the strong Hund-coupling limit, 
spin-wave excitations of a DE ferromagnet below the Curie temperature
$T_C$ can be described by a Heisenberg Hamiltonian with only the
nearest neighbor exchange coupling \cite{furukawa}. At the long 
wavelength (small wavevector $q$), spin-wave stiffness $D$ measures the average
kinetic energy of charge carriers and therefore should increase with
increasing $x$ \cite{furukawa,golosov}.  While spin dynamics of some
manganites initially studied appeared to follow these predictions
\cite{perring96,hirota96}, later measurements revealed anomalous zone
boundary magnon softening deviating from the nearest neighbor
Heisenberg Hamiltonian for other materials with 
$x\sim 0.3$ \cite{hwang98,jaime98,dai00,barilo,chatterji02}.  Three
classes of models have been proposed to explain the origin
of such deviations. The first is based on the DE mechanism,
considering the effect of the on-site Coulomb repulsion \cite{golosov}
or the conducting electron band ($e_g$) filling dependence of the DE
and superexchange interactions \cite{solovyev99}.  The second suggests
that  magnon-phonon coupling \cite{dai00,furukawa99} or the effects of
disorder on the spin excitations of DE systems \cite{motome05} is the
origin for the zone boundary magnon softening. Finally, quantum
fluctuations of the planar $(x^2-y^2)$-type orbital associated with
the A-type antiferromagnetic (AF) ordering may induce magnon softening
as the precursor of such AF order \cite{khaliullin00}.  Although all
these models appear to be reasonable in explaining the zone boundary
magnon softening near $x=0.3$, the lack of complete spin-wave
dispersion data for $A_{1-x}A^{\prime}_{x}$MnO$_3$ with $x<0.3$ and
$x>0.4$ means that one cannot test the doping dependence of different 
mechanisms and, therefore, the origin of the magnon
softening is still unsettled. 

Very recently, Endoh {\it et al.} \cite{endoh05} measured spin-wave
excitations in the FM phase of Sm$_{0.55}$Sr$_{0.45}$MnO$_3$
(SSMO45) and found that the dispersion can be described phenomenologically by the
Heisenberg model with the nearest neighbor ($J_1$) and fourth-nearest
($J_4$) neighbor exchange coupling (Fig.~1). By comparing the 
$J_4/J_1$ ratios, which measure the magnitude of the zone boundary
magnon softening, of SSMO45 with that of 
Pr$_{0.63}$Sr$_{0.37}$MnO$_3$ (PSMO37)  \cite{hwang98} and
La$_{1-x}$Sr$_x$MnO$_3$ ($x=0.2$,0.3; LSMO20, LSMO30)
\cite{hirota96}, the authors concluded that $J_4/J_1$ increases
dramatically for $A_{1-x}A^{\prime}_{x}$MnO$_3$ with
$x>0.3$. Since theoretical analysis based on the local density
approximation $+$ Hubbard U band calculations reveal that this
doping dependence is consistent with the effect of rod-like $(3z^2-r^2)$ 
orbital correlations, the authors
argue that the observed zone boundary magnon softening in
$A_{1-x}A^{\prime}_{x}$MnO$_3$ is due to the
$(3z^2-r^2)$-type orbital fluctuations, in sharp contrast to all
previous proposals \cite{endoh05}.

\begin{figure}[ht!]
\includegraphics[width=2.9in]{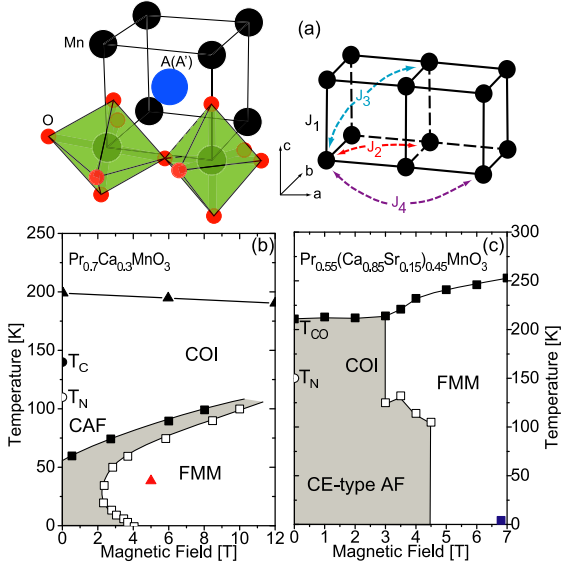}
\caption{\label{fig:fig1} 
(a) Crystal structure of $A_{1-x}A^{\prime}_{x}$MnO$_3$ with magnetic
exchange coupling indicated.  (b) Phase diagram of $\rm Pr_{0.7}Ca_{0.3}MnO_3$ in the
$T$-$H$ plane \cite{tomioka96}.  (c) Phase diagram of $\rm Pr_{0.55}(Ca_{0.85}Sr_{0.15})_{0.45}MnO_3$ in the
$T$-$H$ plane from transport measurements. Our neutron experimental
conditions are marked as the (red) upper triangle and the (blue)
square in the phase diagrams.} 
\end{figure}

In this Letter, we take an approach different from that of Endoh
{\it et al.}, as we know that $J_4/J_1$ is nonzero for $x=0.3$
manganites such as LSMO30 (Figs.~3 and 4) \cite{doloc} and
La$_{0.7}$Ca$_{0.3}$MnO$_3$ (LCMO30) \cite{dai00}, in contrast with
the expectation of Refs. \cite{hirota96,endoh05}. We decided to
systematically analyze all existing spin-wave excitation data and take
additional data in the FM metallic state of
$A_{1-x}A^{\prime}_{x}$MnO$_3$ at judicially selected $x$. We find
that the low-$q$ spin-wave stiffness $D$ is insensitive to $x$ while
spin-wave excitations are systematically renormalized near the zone boundary
with $J_4/J_1$ proportional to $x$. We also find 
 on-site disorder and lattice distortion that control $T_C$'s of
different $x=0.3$ manganites \cite{hwang95} have no effect on $J_4/J_1$,
different from the expectations of the disorder effect theory
\cite{motome05}.  These observations cannot be explained
consistently by any current theory, thus suggesting that more than one
mechanism is at play in determining the spin dynamical properties of
the $A_{1-x}A^{\prime}_{x}$MnO$_3$ manganites.

For this study, we used single crystals of $\rm
La_{0.75}Ca_{0.25}MnO_3$ (LCMO25), $\rm Pr_{0.7}Ca_{0.3}MnO_3$
(PCMO30) and $\rm Pr_{0.55}(Ca_{0.85}Sr_{0.15})_{0.45}MnO_3$
(PCSMO45) grown by the traveling solvent floating zone technique.  We chose these
samples because they represent a large span in carrier
concentrations. While LCMO25 has a FM metallic ground state with
$T_C=191$~K \cite{dai01}, PCMO30 \cite{tomioka96,fernandez02} and
PCSMO45 \cite{tomioka02} exhibit AF insulating
behavior at zero field but can be transformed into FM metallic
phases by field cooling from room temperature (Fig.~1).  Our neutron
scattering experiments were performed on triple-axis spectrometers
at the High-Flux Isotope Reactor (HFIR), Oak Ridge National
Laboratory and the NIST Center for Neutron Research (NCNR). The
momentum transfers $\vec{q}=(q_x, q_y, q_z)$ in units of \AA$^{-1}$
are at positions $(h,k,l)=(q_x a/2\pi,q_y b/2\pi,q_z c/2\pi)$ in
reciprocal lattice units (rlu), where lattice parameters for LCMO25,
PCMO30, and PCSMO45 are $a\approx b\approx c\approx 3.87$~\AA,
3.856~\AA, and 3.834~\AA, respectively.

\begin{figure}[ht!]
\includegraphics[width=3.3in]{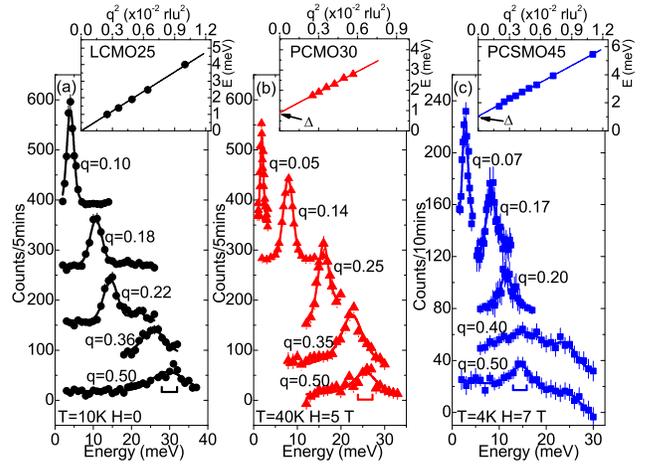}
\caption{\label{fig:fig2} 
The $q$-dependent spin-wave excitations in LCMO25 (a), PCMO30
(b), and PCSMO45 (c) at low temperatures. The data at different $q$'s
are incrementally shifted for clarity. The 
horizontal bars are the instrumental resolution and small shoulders around 25 meV
in (c) are phonon scattering. $E$ versus $q^2$ are plotted in the insets to determine
$D$.}
\end{figure}

Figure 2 shows constant-$q$ scans at representative wavevectors for
LCMO25, PCMO30 and PCSMO45 along the $[1,0,0]$ direction.  
The excitation peaks are sharp and
resolution-limited at low-$q$, but become weak in intensity 
near the zone boundary. To obtain the strength
of the average magnetic interaction, we analyze the low-$q$ data
using $E=\Delta+D q^2$, where $E$ is the spin-wave energy obtained
from Gaussian fits, $\Delta$ is field-induced Zeeman gap
\cite{fernandez02} and $D$ is the spin-wave stiffness.  The slopes of
the $E$ versus $q^2$ lines, shown in the insets of Fig.~2, yield $D$
values of $150\pm3$, $145\pm8$, $152\pm3$ $\rm meV\AA^2$ for LCMO25,
PCMO30 and PCSMO45, respectively.  It is remarkable that all three
samples exhibit very similar low-$q$ behavior after they are driven
into FM states either by temperature or by magnetic field. This
suggests that the average kinetic energy derived from the hopping of
the itinerant electrons between adjacent manganese ions is independent
of carrier concentration.

\begin{figure}[ht!]
\includegraphics[width=2.4in]{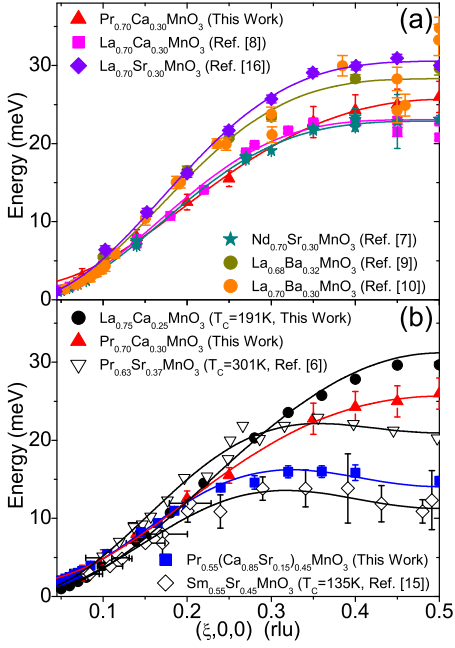}
\caption{\label{fig:fig3}
(a) Spin-wave dispersion curves of various
$A_{0.7}A^{\prime}_{0.3}$MnO$_3$ manganites along the $[\xi,0,0]$
direction. (b) Dispersion curves for a series of
$A_{1-x}A^{\prime}_{x}$MnO$_3$ as a function of $x$.  The solid
lines are least-square fits using the Heisenberg model with $J_1$
and $J_4$. } 
\end{figure}

To determine the evolution of magnetic excitations in
$A_{1-x}A^{\prime}_{x}$MnO$_3$ as a function of $x$, one must first
understand the effect of on-site disorder arising from the mismatch
between rare- and alkaline-earth ions, as such disorder might induce
anomalous spin dynamical behavior \cite{motome05}.  The disorder is
characterized by $ \sigma^2=\sum_i (x_i r_i^2 - {\bar r}^2)$, where
$x_i$ is the fractional occupancies of $A$-site species, $r_i$ and
$\bar r=\sum_i x_i r_i$ are individual and averaged ionic radius
respectively \cite{martinez96,ionsize}. Fig.~3(a) summarizes
spin-wave dispersions along the $[1,0,0]$ direction for a series
of $A_{1-x}A^{\prime}_{x}$MnO$_3$ with $x\approx 0.30$
\cite{jaime98,dai00,barilo,chatterji02,doloc}, while the doping
dependence of magnon excitations is shown in Fig.~3(b)
\cite{hwang98,endoh05}.  The solid lines in the figure are 
phenomenological fits to
the data using Heisenberg Hamiltonian $E(\vec{q}) = \Delta
+2S[J(0)-\sum_{j}J_{ij}e^{i \vec{q} \cdot (\vec{R_i}-\vec{R_j})}]$
with nearest neighbor ($J_1$) and fourth-nearest
neighbor ($J_4$) exchange coupling. In the low-$q$ limit, 
$E(q) = \Delta +8\pi^2 S(J_1 +4 J_4) q^2$.
This simple Hamiltonian gives a satisfactory description of the data,
where $J_4/J_1$ measures the magnitude of the effective zone
boundary magnon softening \cite{endoh05}. We note that our
previous experience \cite{hwang98} showed that $J_2$ and $J_3$ have
small contributions to the total magnon dispersion. 

We are now in a position to determine the effect of on-site disorder
($\sigma^2$) and average ionic size ($\bar r$) on spin-wave
excitations of $A_{0.7}A^{\prime}_{0.3}$MnO$_3$.  Figures 4(a-c) show
the $\sigma^2$ dependence of the spin-wave stiffness $D$, the
nearest neighbor exchange coupling $J_1$ and the ratio of $J_4/J_1$ in
$A_{0.7}A^{\prime}_{0.3}$MnO$_3$. With increasing disorder, the
long-wavelength limit spin-wave stiffness shows no systematic trend
but falls within a bandwidth of $D=160\pm 15$ meV\AA$^2$ [Fig.~4(a)].
While such behavior at low-$q$ is not unexpected \cite{jaime98},
numerical calculations suggest a significant zone boundary magnon
softening with increasing disorder \cite{motome05}. In other words,
increasing $\sigma^2$ should have no effect on $D$ but dramatically
increase $J_4/J_1$. Surprisingly, Figure 3(a) reveals no direct
correlation between $\sigma^2$ and zone boundary magnon energy; and
Figure 4(c) shows that $J_4/J_1$ is independent of $\sigma^2$.
Therefore, the on-site disorder has no observable effect on zone
boundary magnon softening. 

\begin{figure}[ht!]
\includegraphics[width=3.2in]{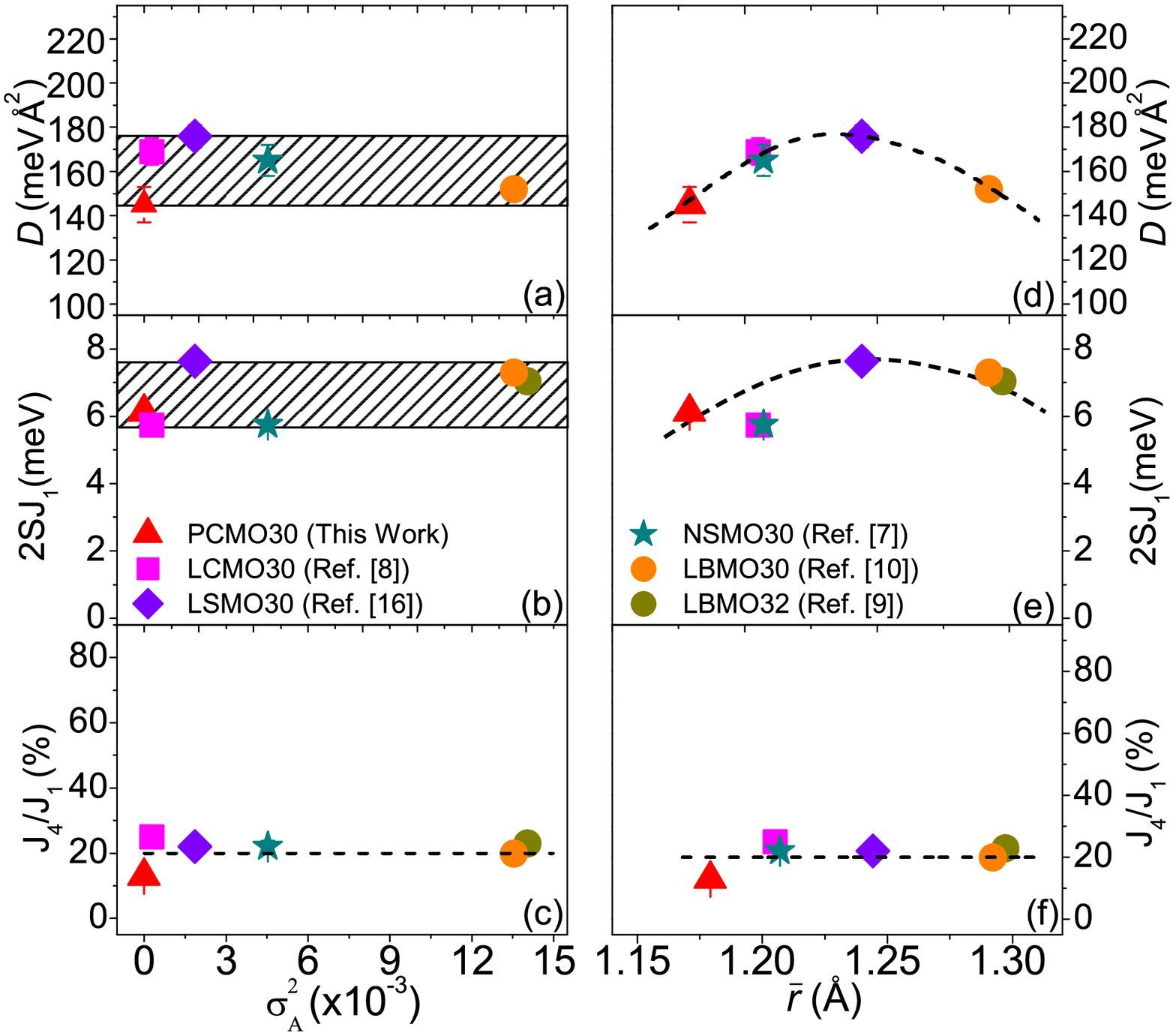}
\caption{\label{fig:fig4}
The on-site disorder dependence of (a) the spin-wave stiffness $D$
obtained from low-$q$ excitations, (b) the coupling $2SJ_1$ and (c)
the ratio of $J_4/J_1$. The average ionic radius dependence of (d)
$D$, (e) $2SJ_1$, and (f) $J_4/J_1$. Dashed lines are guides to the
eye.
} 
\end{figure}

In addition to inducing on-site disorder, replacing $A$ by
$A^\prime$ in $A_{0.7}A^{\prime}_{0.3}$MnO$_3$ will also change the
average ionic radius $\bar r$ and modify the length and angle of
Mn-O-Mn bonds, thus leading to changes in effective transfer
integrals between Mn ions or the bandwidth of the electrons
\cite{hwang95}.   Figures 4(d-f) show the $\bar r$ dependence of
the spin-wave stiffness $D$, $2SJ_1$, and $J_4/J_1$. With
increasing $\bar r$, $D$ shows a parabolic curve within a small
bandwidth. It increases from 160 meV\AA$^2$ at $\bar r\sim 1.21$
\AA\ for NSMO30 to 176 meV\AA$^2$ at $1.25$ \AA\ for LSMO30, and
then decreases to 152 meV\AA$^2$ at 1.29 \AA\ for LBMO30
[Fig.~4(d)]. The $T_C$'s for NSMO30, LSMO30, and LBMO30 are 198 K
\cite{jaime98}, 351 K \cite{doloc}, and 350 K \cite{chatterji02}
respectively. Although increasing $\bar r$ leads to rapid changes
in $T_C$, the kinetic energy ($D$) or bandwidth of the
electrons only changes slightly \cite{jaime98,radaelli97}.
Furthermore, $J_4/J_1$ is independent of $\bar r$ [Fig.~4(f)], thus
indicating that the magnitude of the zone boundary magnon softening is
independent of $T_C$ and a general feature of the
$A_{0.7}A^{\prime}_{0.3}$MnO$_3$ manganites.

\begin{figure}[ht!]
\includegraphics[width=2.6in]{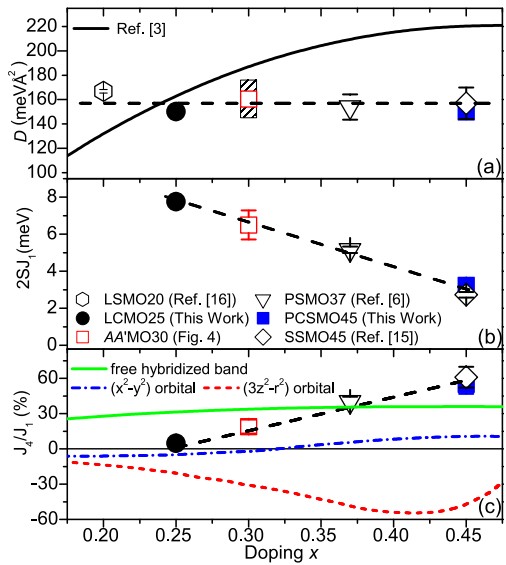}
\caption{\label{fig:fig5}
Doping dependence of (a) spin-wave stiffness $D$, (b) nearest
neighbor exchange coupling $J_1$ and (c) ratio of $J_4/J_1$.
Dashed lines are guide to the eye. The solid line in (a) is the
prediction of Ref.~[3]. The (green) solid, (blue) dash-dotted, and
(red) dashed lines in (c) are calculations from different models in
\cite{endoh05}.
} 
\end{figure}

Assuming the effect of on-site disorder and ionic size is weakly
doping dependent, we can then study how spin-wave excitations of
$A_{1-x}A^{\prime}_{x}$MnO$_3$ are modified as a function of $x$.  
For LSMO20, we used the stiffness value of
$D=166.8\pm 1.51$ meV\AA$^2$ obtained by high-resolution cold neutron triple-axis at
low-$q$ \cite{doloc} because this value is more accurate than the earlier result
of 120 meV\AA$^2$ obtained on a thermal triple-axis \cite{hirota96}.
Figure 5 summarizes the $x$ dependence of $D$, $2SJ_1$, and
$J_4/J_1$. Surprisingly, the spin-wave stiffness $D$ is around $160\pm
15$ meV\AA$^2$ and essentially unchanged for $0.2\leq x\leq 0.45$
[Fig. 5(a)], while $T_C$ varies from 305 K for LSMO20 \cite{doloc} to
135 K for SSMO45 \cite{endoh05}.
This is in sharp contrast to the expectation of all DE based models,
where $D$ increases with $x$ [solid line in Fig.~5(a)] \cite{golosov}.
This also differs from a conventional ferromagnet, where $T_C$ should
be proportional to $D$.  On the other hand, $2SJ_1$ decreases and
$J_4/J_1$ increases, approximately linearly, with increasing $x$
[dashed lines in Figs.~5(b,c)].  Therefore, $J_4/J_1$ does not exhibit
a huge rise in magnitude  for $x\geq 0.4$ as expected from the
$(3z^2-r^2)$-type orbital fluctuations (dashed line), nor does it follow
the predictions of the $(x^2-y^2)$-type orbital fluctuations (dash-dotted
line) or free hybridized bands (solid line) shown in Fig.~5(c)
\cite{endoh05}.  

Our systematic investigations in Figs.~4 and 5 put stringent
constraints on microscopic theories of the zone boundary magnon
softening.  The possibility of on-site disorder-induced zone
boundary softening is ruled out, as such a theory expects an enhanced
softening with either increasing disorder or decreasing $x$
\cite{motome05}, both contrary to the observation. Similarly, it is
unclear how on-site Coulomb repulsion in a DE mechanism can explain
the doping independent behavior of the spin-wave stiffness
\cite{golosov}. The magnon and $A_{1-x}A^{\prime}_x$-site optical
phonon coupling (crossing) scenario postulated for LCMO30
\cite{dai00} also has difficulty in explaining the evolution of
$J_4/J_1$, as the average $A_{1-x}A^{\prime}_x$-site mass and
frequencies of associated optical phonon modes do not vary
dramatically from LCMO25 to SSMO45. If the large $J_4/J_1$ for
$x=0.45$ materials stems from fluctuations of the $(3z^2-r^2)$
orbital, $J_4/J_1$ should have a spectacular doping dependence around
$x=0.4$ \cite{endoh05}. However, this is not observed. Furthermore,
the $(3z^2-r^2)$-type orbital fluctuations should have little or no
effect on spin-wave softening at low carrier doping of $x=0.25$ and
$0.3$.  Therefore, it cannot be the origin of zone boundary magnon
softening in $A_{1-x}A^{\prime}_{x}$MnO$_3$ at all doping levels.
Finally, we note that the free hybridized band model and $(x^2-y^2)$
orbital fluctuation effects do not have the correct doping dependence
for $J_4/J_1$ [Fig.~5(c)]. Although none of the current theories is
capable of explaining the spin dynamics in the entire doping regime,
it is possible that more than one effect determines the properties of
spin excitations in manganites. For example, strong Coulomb repulsion
in connection with the electron-phonon coupling can induce orbital
polarization throughout the phase diagram. It is also known that
chemical disorder can fundamentally modify the ground states of
manganites \cite{akhoshi} and be responsible for changes in spin
dynamics \cite{sato}. To understand the evolution of spin dynamical
behavior in $A_{1-x}A^{\prime}_{x}$MnO$_3$, one must consider
interactions among spin, charge, orbital, and lattice degrees of
freedom. 

This work is supported by the U.S. NSF DMR-0453804,
DOE Nos. DE-FG02-05ER46202 and DE-AC05-00OR22725 with UT/Bettelle LLC, 
and also by NSF DMR-0346826 and 
DOE DE-FG02-05ER46125. This work was performed under the US-Japan Cooperative
Program on Neutron Scattering.

\end{document}